\documentclass{aa} 

\usepackage{graphicx}
\usepackage{color}
\usepackage{txfonts}
\usepackage{float}
\begin{document}

   \title{Interpreting observations of edge-on gravitationally unstable accretion flows}

   \subtitle{The case of G10.6-0.4}

   \author{Hauyu Baobab Liu\inst{1}
          }

   \institute{European Southern Observatory (ESO), Karl-Schwarzschild-Str. 2, D-85748 Garching, Germany \\
                 \email{baobabyoo@gmail.com}
             }

   \date{Received August 25, 2016; accepted September 21, 2017}

  \abstract
   {Gravitational collapse of molecular cloud or cloud core/clump may lead to the formation of geometrically flattened, rotating accretion flow surrounding the new born star or star cluster. Gravitational instability may occur in such accretion flow when the gas to stellar mass ratio is high (e.g. over $\sim$10\%).}
   {This paper takes the OB cluster-forming region G10.6-0.4 as an example. We introduce the enclosed gas mass around its central ultra compact (UC) H\textsc{ii} region, addresses the gravitational stability of the accreting gas, and outline the observed potential signatures of gravitational instability.}
   {The dense gas accretion flow around the central UC H\textsc{ii} region in G10.6-0.4 is geometrically flattened, and is approximately in an edge-on projection. The position-velocity (PV) diagrams of various molecular gas tracers on G10.6-0.4 consistently show asymmetry in the spatial and the velocity domain. We deduce the morphology of the dense gas accretion flow by modeling velocity distribution of the azimuthally asymmetric gas structures, and by directly de-projecting the PV diagrams.}
   {We found that within the 0.3 pc radius, an infall velocity of 1-2 km\,s$^{-1}$ may be required to explain the observed PV diagrams. In addition, the velocity distribution traced in the PV diagrams can be interpreted by spiral arm-like structures, which may be connected with exterior infalling gas filaments. The morphology of dense gas structures we propose appears very similar to the spatially resolved gas structures around the OB cluster-forming region G33.92+0.11 with similar gas mass and size, which however is likely to be approximately in a face-on projection.} 
   {The dense gas accretion flow around G10.6-0.4 appears to be Toomre unstable, which is consistent with the existence of large-scale spiral arm-like structures, and the formation of localize gas condensations. The proposed approaches for data analyses may be applied to the observations of Class 0/I low-mass protostars, to diagnose disk gravitational instability.}


\titlerunning{Gravitational instability at the convergence of accreting molecular gas stream}

   \maketitle
   
\section{Introduction}
Dense gas in star- or cluster-forming molecular cores/clumps may first condense onto a flattened rotating accretion flow or disk, before further accrete onto the (proto)stars. 
When the mass infall rate onto the flattened rotating accretion flow or disk is higher than the (proto)stellar accretion rate, the accumulated dense gas may trigger gravitational instability, leading to the formation of spiral arm-like gas structures and self-gravitating gas condensations (e.g. Vorobyov 2013; Vorobyov \& Basu 2015; Dong et al. 2016).
Such gravitational instability may occur in the Class 0/I stage of low-mass protostars, when the protostellar masses and the stellar-to-disk mass ratios are small.
The accretion disk gravitational instability may resolve the well-known {\it luminosity problem} of the low-mass protostars, and may explain the episodic protostellar accretion and the formation of wide-orbit massive planets (Machida et al. 2011; Dunham \& Vorobyov 2012; Liu et al. 2016). 
The similar phenomenon may be expected from the dense gas accretion flow surrounding the young OB stars (e.g. Sakurai et al. 2016).
However, OB stars may disperse the accreting gas via radiative feedback, before the majority of dense gas can be accreted onto the (proto)stars or Keplerian rotating disks.

\begin{figure*}
\hspace{0.25cm}
\includegraphics[width=18cm]{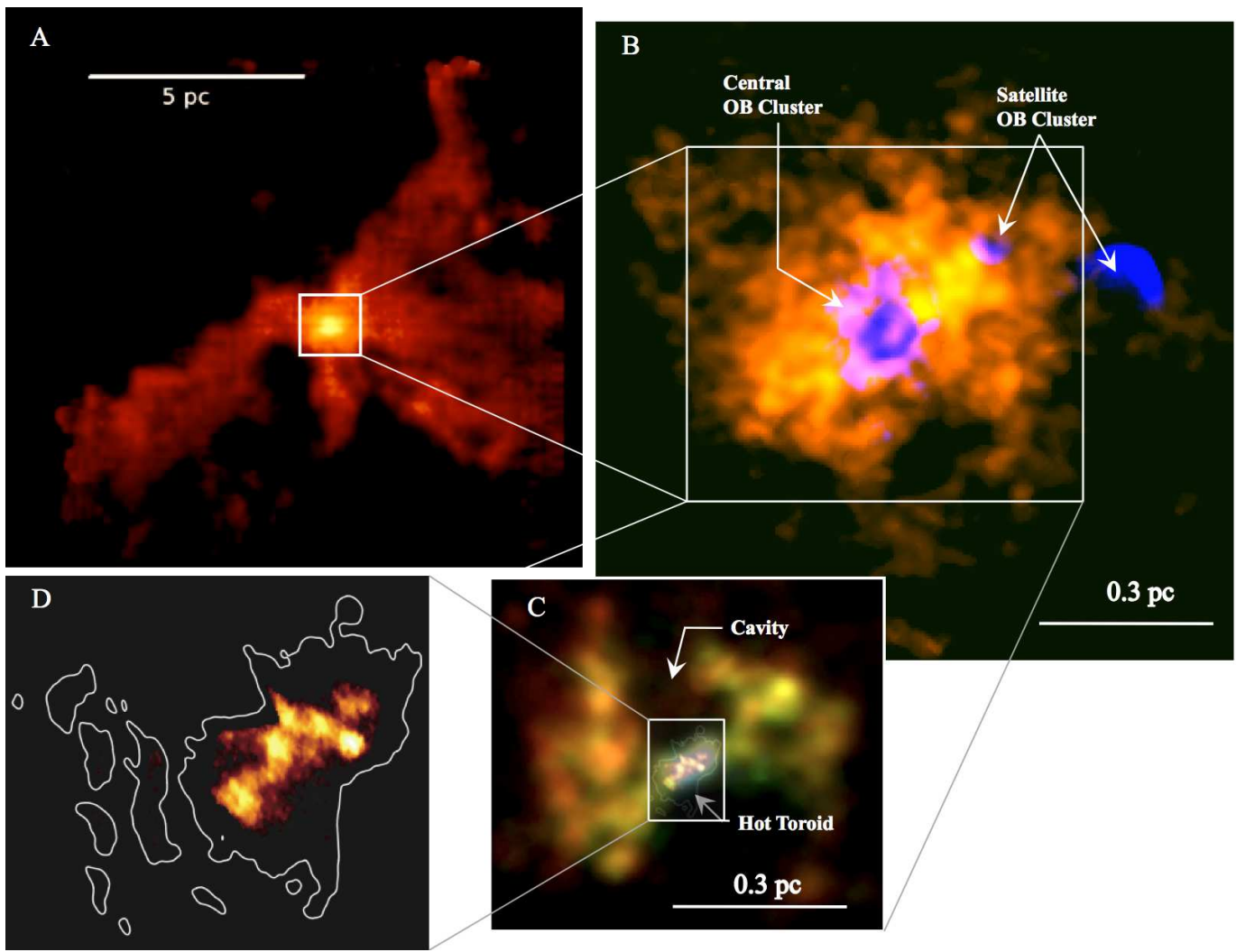}
\caption{\footnotesize{
Gas structures in OB cluster-forming region G10.6-0.4. (A) IRAM-30m telescope observations of $^{13}$CO 2-1, integrated from $-$3 to 0.6 km\,s$^{-1}$ (Liu et al. in prep.) We omit integrating the blueshifted component in this panel to avoid confusing the gas structures due to blending. (B) Color composite image made from the VLA observations of the CS 1-0 line (orange) and the 3.6 cm continuum emission (blue), which were presented in Liu et al. (2011). The deficit of CS 1-0 emission at the center is due to absorption line against the bright UC H\textsc{ii} region. (C) Color composite image made from the SMA observations of three CH$_{3}$OH transitions, which trace 35 K (red), 60 K (green), and 97 K (blue) upper level energy, respectively. Inset overlays the VLA observations of the optical depth map of the NH$_{3}$ (3,3) satellite hyperfine line, which was reproduced from Sollins \& Ho (2005). (D) VLA (A-array configuration) image of NH$_{3}$ (3,3) satellite hyperfine line optical depth map (Sollins \& Ho 2005).
}
}
\label{fig:overview}
\end{figure*}

Molecular cloud G10.6-0.4 (R.A.: 18$^{\mbox{h}}$10$^{\mbox{m}}$28$^{\mbox{s}}$.683, Decl.: $-$19$^{\circ}$55$'$49$''$.07) is a luminous ($\sim$10$^{6}$ L$_{\odot}$) OB cluster forming region at a distance of $\sim$6 kpc\footnote{Uncertainty of distance may be $\sim$20\%.}.
The previous high angular resolution mapping survey of Lin et al. (2016) for $L\ge$10$^{6}$ $L_{\odot}$ star-forming molecular clouds suggested that G10.6-0.4 has an exceptionally centrally concentrated gas distribution, which may be conducive to a focused global cloud collapse.
On the 5-10 pc scale, this cloud presents several approximately radially aligned dense gas filaments, which connect to the central $\sim$1 pc scale massive molecular clump (Figure \ref{fig:overview}A; see Liu et al. 2012a for more details).
The higher angular resolution Very Large Array (VLA) and Submillimeter Array (SMA) observations of dense molecular gas tracers found that some large-scale filaments may converge to a rotating, $\sim$0.6 pc scale massive molecular envelope (Figure \ref{fig:overview}B, \ref{fig:overview}C; Omodaka et al. 1992; Ho et al. 1994; Liu et al. 2010a, 2010b, 2011).
The most massive OB stars, and their associated ultra compact (UC) H\textsc{ii} region (Figure \ref{fig:overview}B; Ho et al. 1986; Ho \& Haschick 1986; Sollins et al. 2005) are deeply embedded in this rotationally flattened (Keto, Ho \& Haschick 1987, 1988; Guilloteau et al. 1988) massive molecular envelope, where the low density gas around the rotational axis has been photo-ionized, or were swept away by expanding ionized gas (Figure \ref{fig:overview}C; Liu et al. 2010b, 2011).
The observations of high excitation molecular lines exclusively traced a $\sim$0.1 pc scale, $\gtrsim$100 K hot molecular toroid at the center of the massive molecular envelope (Liu et al. 2010b; Beltran et al. 2011), which harbors the luminous central OB cluster (Figure \ref{fig:overview}B, \ref{fig:overview}C).
The 0$\farcs$1 resolution VLA observations of the NH$_{3}$ (3,3) satellite hyperfine line absorption shows that this hot toroid is extremely clumpy (Figure \ref{fig:overview}C, \ref{fig:overview}D; Sollins \& Ho 2005). 
The overall geometric configuration of the central $\sim$1 pc scale massive molecular clump resembles a scaled-up, low-mass star-forming core+disk system viewed edge-on.
How the dense gas condensations in the hot toroid came into the existence remains uncertain.
A intriguing possibility is the formation of the dense condensations due to gravitational instability developed in the mid-plane of the flattened rotating massive molecular envelope.
If this is indeed the case, we expect the distribution of dense gas in the flattened accretion flow to present a significant azimuthal asymmetry. 


Wihin the 0.1-0.3 pc radii, the previous observations, limited by angular resolution or sensitivity, did not yet spatially resolve azimuthal asymmetry of dense gas structures. 
Nevertheless, it is possible to diagnose spatial asymmetry base on the velocity profiles of dense molecular gas tracers, which is analogous to the common applications of the optical and infrared spectroscopic observations towards low-mass protostars (e.g. Takami et al. 2016).
In particular, the earlier centimeter band observations of molecular line absorption against the central UC H\textsc{ii} region have suggested a {\it red excess} in the velocity field, which was interpreted as a signature of redshifted infall (Ho \& Haschick 1986; Keto et al. 1987, 1988; Keto 1990; Sollins \& Ho 2005).
The follow-up interferometric observations of multiple molecular line tracers in the millimeter band have confirmed the red excess in the central $\sim$0.06 pc ($\sim$2$''$) region, however, did not detect the blueshifted counterpart of the infalling gas (Liu et al. 2010a, 2011).
The dense molecular gas immediately around the central UC H\textsc{ii} region appears to have an azimuthally asymmetric distribution.
In addition, the position-velocity (PV) diagrams of those molecular lines suggest that the non-axisymmetric dense gas structures can be extended to $\pm$5$''$-10$''$ (0.15-0.3 pc) radii.

This paper elaborates two techniques for diagnosing the asymmetry of dense gas structures based on the spatially resolved line spectra.
In addition, we propose a link between the spatial asymmetry and the gravitational instability.
Although we take the case of the OB cluster-forming region G10.6-0.4 as an example, we argue that the proposed strategy of data analysis/interpretation is general for both low-mass and high-mass star- or cluster-forming regions.
Our discussion begin with a toy model, which describes the major features in the observed molecular line PV diagrams in Section \ref{sec:toy}.
Section \ref{sec:deproject} discusses the probable dense gas morphology based on direct de-projections.
Section \ref{sec:toomre} analyzes the enclosed mass profile around the central UC H\textsc{ii} region, and derives the Toomre Q parameter as a function of radius. 
In section \ref{sec:discussion}, we compare G10.6-0.4 with a spatially resolved example G33.92+0.11, which is likely to be in close to face-on projection.


\begin{figure}
\hspace{-0.3cm}
\includegraphics[width=9.2cm]{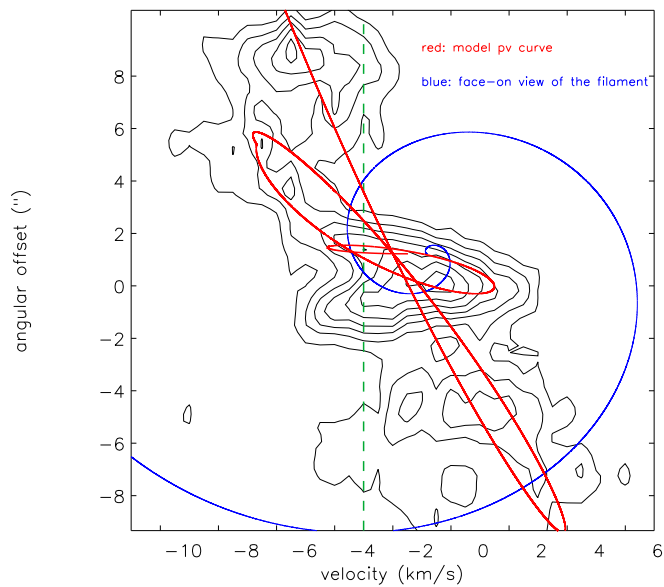}
\vspace{-1.3cm}
\caption{\footnotesize{
Position-velocity diagram of the CH$_{3}$OH 5$_{0,5}$--4$_{0,4}$ E transition in OB cluster forming region G10.6-0.4 (contour; see Figure 3 of Liu et al. 2011 for the comparisons with other CH$_{3}$OH excitation levels and the NH$_{3}$ (3,3) hyperfine lines), and the model PV curve (red curve). The pv cut is centered at the coordinates of R.A. = 18$^{\mbox{h}}$10$^{\mbox{m}}$28.64$^{\mbox{s}}$ and Decl = $-$19$^{\circ}$55$'$49.22$''$ with position angle pa = 140$^{\circ}$.
The positive angular offset is defined in southeast. 
Contour levels are 0.36 Jy beam$^{-1}$$\times$[1, 2, 3, 4, 5, 6, 7]. 
The resolution of this CH$_{3}$OH observation is 1$''$.5$\times$1$''$.3, and the rms noise level is 0.06 Jy beam$^{-1}$.
The dashed line marks the cloud systemic velocity of $-$3 kms$^{-1}$.
The face-on view of the model filamentary spiral structure is also plotted on the same physical scale (blue curve; assuming observers are viewing this structure from the left, and assuming the same spatial scales in the horizontal and vertical directions).
}}
\label{fig:spiral}
\end{figure}

\section{Toy Model to Interpret the Asymmetry in Position--Velocity Diagram}
\label{sec:toy}
Despite that the dense gas structures in an edge-on source may be blended thus cannot be spatially resolved.
It may still be possible to differentiate gas structures from the spectral domain.
For a three dimension position-position-velocity (PPV) image cube, this can be diagnosed by generating two dimensional PV diagrams. 

Figure \ref{fig:spiral} shows the PV diagram of the CH$_{3}$OH 5$_{0,5}$--4$_{0,4}$ E line ($E_{up}$$\sim$48 K), which trace warm and dense molecular gas immediately around the central OB cluster (Figure \ref{fig:overview}).
The selected pv cut is approximately centered at the central OB cluster in G10.6-0.4 (Figure \ref{fig:overview}B), and is parallel to the plane of the flattened rotating dense gas accretion flow.
On the $\pm$8$''$ scales, the PV diagram traces a negative velocity gradient (i.e. velocity becomes more redshifted as we look at the more negative angular offset).
The blueshifted (relative to the systemic velocity) gas at $\gtrsim$3$''$ angular offset, and the redshifted gas at $\lesssim$$-$3$''$ angular offset, appear very well separated by the $-$3 km\,s$^{-1}$ systemic velocity.
Gas structures within the $\sim\pm$3$''$ angular offset are dominated by a redshifted component with highly negative velocity gradient (hereafter the hot toroid, following the nomenclature in the literature), and a bright emission clump around $-$1$''$ angular offset and $-$3 km\,s$^{-1}$ line-of-sight velocity.
These two components are blended in the PV diagram presented in Figure \ref{fig:spiral}.
They can be separated since the observations of the high excitation CH$_{3}$OH and CH$_{3}$CN lines exclusively trace the hot toroid component (Liu et al. 2010a, 2010b), and since the previous absorption line experiment only detected the hot toroid component (Sollins \& Ho 2005).
The hot toroid components seems to connect to the extended blueshifted gas via a dense ridge extending from 3$''$ to 6$''$ angular offset. 
This dense ridge may be clumpy, which was also shown by the previous observations of the optically thinner, $^{13}$CS 5-4 line (Liu et al. 2010a).

At least part of these gas structures may be explained by spiral arm-like structures which are orbiting the central OB cluster.
Motivated by the proposed spinning-up motion from the previous observations (e.g. Keto et al. 1987; Keto 1990), we attempt spiral arm models which are tightly wound in inner region but quickly become loose at outer radii.
Such (single) spiral arm structure may be simply realized by the following empirical formula:
\begin{equation}
\theta = -(r\cdot m)^{\beta} - \phi_{0},
\label{eq_spiral}
\end{equation}
where $\theta$ is the position angle relative to the reference position angle $\phi_{0}$, $r$ is the radius, $m$ and $\beta$ are parameters to control how the spiral arm is wrapped from small to large radii.

By referencing to the previously reported enclosed mass and the observed rotation curve (Liu et al. 2010), we approximate the absolute value of velocity of gas (i.e. gas speed) as a function of radius, by the following empirical power-law for the sake of simplicity: 
\begin{equation}
v(r) = v_{0}\left(\frac{r}{r_{0}}\right)^{\alpha+1},
\label{eq_v}
\end{equation} 
where $v(r)$ is gas speed at the certain radius $r$; $\alpha$ is observationally constrained to be $-$0.75;  v$_{0}$ and r$_{0}$ are chosen to be 1.75 kms$^{-1}$ and 500 AU to match the observed PV diagram.
Inside the $\sim$0.05 pc radius, this power-law velocity curve underestimates the gas velocity.
Nevertheless, gas within such a small scale is likely mostly photo-ionized (Keto \& Wood 2006), thus does not present highly blueshifted and redshifted emission in the PV diagrams for molecular lines (Figure \ref{fig:spiral}).
Therefore, the assumed power-law velocity curve reasonably well describes the observed molecular gas structures.


For each point on the curve described by Equation \ref{eq_spiral}, we evaluate the line-of-sight velocity based on the gas speed described by Equation \ref{eq_v}. 
We then compare the obtained line-of-sight velocity distribution with the PV diagram of the CH$_{3}$OH  5$_{0,5}$--4$_{0,4}$ E transition (Figure \ref{fig:spiral}).
We make trials for the combinations of rotation and infall velocities. 
The model best fits to the most significant structures in the PV diagram (Figure \ref{fig:spiral}) assumes that the inward radial velocity is about 30\% of the gas speed ($\sim$0.9-1.2 kms$^{-1}$) inside the 0.125 pc radius.
Gas outside of the 0.125 pc radius follows pure rotational motion.
The values of $m$, $\beta$, and $\phi_{0}$ are 1.2$\cdot$10$^{-1}$ ($\frac{1}{[au]}$), 0.295, and 0.0, respectively.
The simple spiral arm model (described by Equation \ref{eq_spiral}) successfully explain the {\it red excess} of the hot toroid velocity field, the connection from the hot toroid to the blueshifted, positive angular offset ridge, and the more extended gas features.

We note that the simple toy model introduced in this section is to provide an insight into the shape of spiral arms in a PV diagram.
One intriguing remark based on this analysis is that, to correctly interpret the gas kinematics, it is necessary to observationally resolve the connection of gas structures from small to large scales, probably from the observations of more than one molecular gas tracers.
For example, if we only observe exclusive tracers of the hot toroid (e.g. the CH$_{3}$CN lines, see Klaassen et al. 2009; Liu et al. 2010a; Beltr\'an et al. 2011) but without any complementary information, then the interpretation of its velocity gradient may be degenerated. 
It may not be easy to distinguish a purely rotating hot toroid at a slightly more redshifted systemic velocity, from the azimuthally asymmetric gas structure with both infall and rotational velocity.
This degeneracy can lead to errors in the estimates of the embedded stellar mass, although does not significantly bias the discussion about gas structures in the present paper since the molecular gas mass is more dominant (more discussion in Section \ref{sec:deproject}).
In addition, exercises in this section re-invoke that the velocity line profiles of a gas flow can be either blue-skewed or red-skewed due to its azimuthal asymmetry rather than infall or outflow motions and self-absorption.
In such cases, diagnosing self-absorption will require observing tracers for different optical depths.

It is not realistic to consider that a single smooth spiral-arm can describe all dense gas structures in the accretion flow.
There may be more spiral-arms or localized dense gas condensations.
For example, the significant emission around $-$3 kms$^{-1}$ and the angular offset $-$1$''$ in the PV diagram (Figure \ref{fig:spiral}) is not described by our best fit single spiral-arm model, however, can be explained by other spiral-arm structures or filamentary accretion flows in the line-of-sight.
This component may be interpreted by the innermost part of the blueshifted infalling molecular gas from the far side, hence was not detected in the centimeter band absorption line experiment (Sollins \& Ho 2005).
The more blueshifted molecular gas is not presented either due to photo-ionization, or may be simply explained by part of the azimuthal asymmetry of the molecular gas accretion flows.
In addition, a spiral-arm can be spatially discontinuous (e.g. due to gravitational instability or stellar feedback), and therefore miss to present strong molecular line emission at certain expected sectors of angular offsets and line-of-sight velocity.
Moreover, there may be relatively diffuse gas components, which can occupy a large line-of-sight velocity range at each angular offset.
Such diffuse components are less trivial to be described with a model. 
A systematic de-projection can be an alternative approach to reconstruct the geometry of the dense gas accretion flow. 

\begin{figure}
\vspace{-5.1cm}
\hspace{-0.3cm}
\includegraphics[width=10cm]{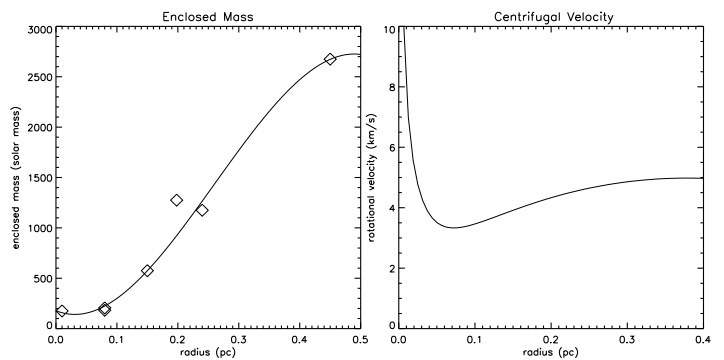}
\caption{\footnotesize{Enclosed stellar and and molecular gas mass $M_{enc}$ as a function of radius $r$, in OB cluster-forming region G10.6-0.4.
Left:-- The $M_{enc}$ at various radii (diamonds) quoted from Liu et al. (2010).
We overplot a 3rd-order polynomial fit to these data.  
Right:-- Absolute value of velocity estimated by $v$$=$$GM_{enc}(r)$/$r$, where $G$ is the gravitational constant.
}}
\label{fig:enclosed}
\end{figure}

\section{De-projected accretion stream structures}\label{sec:deproject}

\begin{figure}
\vspace{-5.1cm}
\includegraphics[width=9.5cm]{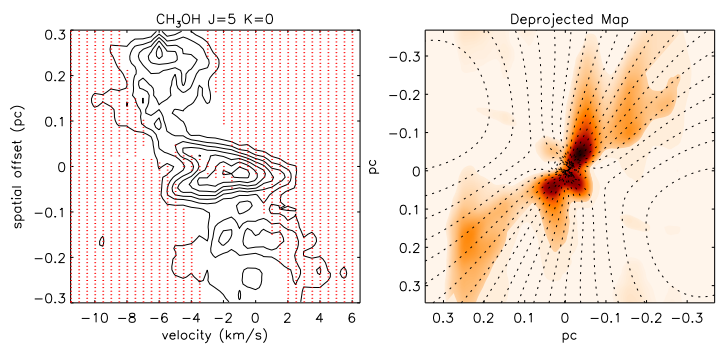}
\caption{\footnotesize{
De-projection analysis of the PV diagram of the OB cluster forming region G10.6-0.4 ($\psi$ = 25$^{\circ}$).
To perform the de-projection, we evaluated the absolute velocity field based on the enclosed mass. 
We adopt the value of centrally embedded stellar mass to be 175 $M_{\odot}$ (Sollins \& Ho 2005).
The velocity vectors are twisted from pure rotational motions toward the center by 25$^{\circ}$.
Left:-- The PV diagram of the CH$_{3}$OH 5$_{0,5}$--4$_{0,4}$ E transition in G10.6-0.4 (see also Liu et al. 2011). 
Contour levels are 0.36 Jy beam$^{-1}$$\times$[1, 2, 3, 4, 5, 6, 7]. 
Regions in this PV diagram that cannot be explained with the velocity field model are masked with red dots.  
Right:-- The 2 dimensional image of the de--projected structures. Dashed lines are the iso line-of-sight velocity contours, start from $-$5.0 km\,s$^{-1}$, separated by $\pm$0.5 km\,s$^{-1}$.
The de-projected image shows elongated large scale structure ($\gtrsim$0.5 pc) from bottom left to top right.
More complicated structures are seen within the central $\sim$0.15 pc radius.
}
}
\label{fig_deproject1}
\end{figure}

\begin{figure}
\vspace{-4.5cm}
\includegraphics[width=9cm]{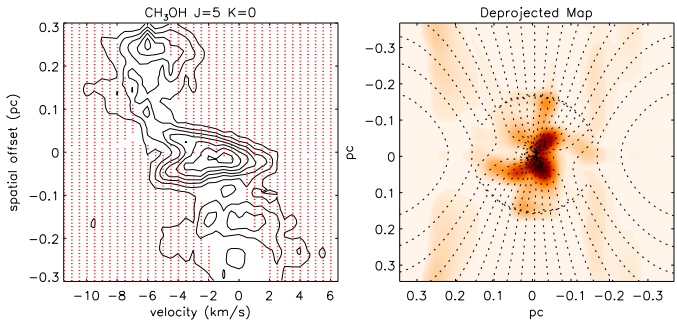}
\caption{\footnotesize{
Similar with Figure \ref{fig_deproject1}. However, we adopt a rotational velocity dominated model ($\psi$ = 0$^{\circ}$ for $r$$>$0.16 pc; $\psi$ = 15$^{\circ}$ for $r$$<$0.16 pc). 
In this model, the motions outside of the 0.16 pc radius is purely rotational. 
Inward of the 0.16 pc radius, the velocity vectors are twisted toward the center by 15$^{\circ}$. 
This gives some infall velocity in the inner regions that is necessary to explain the PV diagram. 
Besides, adding little infall velocity suppress the velocity degeneracy during the de-projection.
The de--projected image shows smoothed large scale structure ($\gtrsim$0.5 pc), and shows molecular arms within the central $\sim$0.15 pc radius.
}
}
\label{fig_deproject2}
\end{figure}

The single component model in the previous section helps gain insight into interpretation of the asymmetry in the PV diagram. 
For a realistic system, the single component model is not necessarily sufficient to incorporate all structures. 
This problem can be alleviated by introducing more components in the model, nevertheless, damages the objectivity. 
Alternatively, the observed line-of-sight velocity $v_{lsr}$ at certain coordinates (e.g. R.A., Decl.) may be described by projecting a three dimensional velocity model $v(x,y,z)$.
For example, assuming that the most significant structures in the PV diagram of G10.6-0.4 are moving in a flattened plane of rotation, we can invert the velocity mapping, to find the de-projected line-of-sight locations, for gas components at certain coordinates and $v_{lsr}$. 
The observed PV diagrams may be systematically de-projected into two dimensional matter distribution based on a priori assumption of velocity field.  
Whether or not the adopted model of velocity field is plausible may be checked by comparing the expected distribution of the line of sight velocity with the PV diagram. 

The primary assumptions we use in the de-projection are:
\begin{itemize}
\item[1.] The most massive stars (e.g. $\ge$5 $M_{\odot}$) are concentrated to the central $\ll$0.1 pc region, and are embedded in the UC H\textsc{ii} region.
\item[2.] Though the molecular gas does not have an azimuthally symmetric distribution, the motions of the molecular gas are described by an azimuthally symmetric velocity field. 
\item[3.] The most significant observed structures only have planar, infall or rotational motions.   
\end{itemize}

We adopt the systemic velocity of the molecular cloud G10.6-0.4 to be $-$3 km\,s$^{-1}$, which is the average velocity from the large-scale cloud structures, and was commonly used in literature.
The assumption 1 and 3 can be modified with more accurate observations.
The assumption of the azimuthally symmetric velocity field may be reasonable, if there are sufficient interactions between gas structures that relax the motions from their initial conditions. 

In our simplistic approach, the scale of the velocity field $v(r)$ relative to the center of the system is set to be equal to the rotational speed that needs to balance the gravitational acceleration by the embedded mass. 
Values of the enclosed molecular gas mass in certain radii are quoted from Liu et al. (2010).
We use a third order polynomial fitting to interpolate the embedded mass in between radii that have observational measurements. 
We assume an embedded stellar mass of 175 $M_{\odot}$ (Sollins \& Ho 2005) or 69 $M_{\odot}$ (Klaassen et al. 2009) at the center of the system.
Results of de-projection based on the two different assumptions of embedded stellar masses are compared.
Figure \ref{fig:enclosed} shows the embedded stellar and molecular gas mass, and the corresponding rotational speed (for the case of 175 $M_{\odot}$ embedded stellar mass), as functions of radius.
Practically, we assign the velocity vector a position angle $\psi$ relative to the azimuthal direction, to realize both the infall and the rotational motion.
We note that here we make assumptions about the velocity field in an empirical way rather than adopting any (semi-)analytic formulation, since it is not trivial to generalize (semi-)analytic formulations for the cases that the enclosed molecular gas mass dominates the overall embedded mass.

Left panel of Figure \ref{fig_deproject1} compares the expected line of sight velocity assuming the constant value $\psi$=25$^{\circ}$ across all radii, with the CH$_{3}$OH  5$_{0,5}$--4$_{0,4}$ E PV diagram of G10.6-0.4. 
This $\psi$ value gives infall velocity $v_{in}(r)$ to be $\sin$($\psi$)$\sim$0.422 times $v(r)$; and the rotational velocity $v_{rot}$ to be $\cos$($\psi$)$\sim$0.906 times $v(r)$.
We found that the velocity model with this simplistic assumption of $\psi$ reasonably fits the observed features in the PV diagram. 
For the molecular gas distributed in between 0.1-0.3 pc radii, this $\psi$ value implies an infall velocity of 1.6$\pm$0.2 km\,s$^{-1}$. 
The rotational velocity of the molecular gas distributed in between the radius of 0.1 and 0.3 pc is slowly decreasing, from 4.1 to 3.2 km\,s$^{-1}$.
This rotational velocity field is consistent with the terminal velocity analysis in Liu et al. (2010), which suggests a rapid decrease of the specific angular momentum from 1.23 km\,s$^{-1}$\,pc to 0.32 km\,s$^{-1}$\,pc from outer to inner radii.
The right panel of Figure \ref{fig_deproject1} shows the de-projected, two dimensional spatial distribution of molecular gas. 
Whenever the de-projection is subject to positional degeneracy (i.e. if the same line-of-sight velocity can appear more than once along one line-of-sight), our program distributed equal amount of emission to all possible line-of-sight positions.
The spatial resolution of this de-projected image is limited by linewidth of the observed molecular transitions, and the velocity resolution of the observations. 
The de-projected structures are artificially stretched when the iso velocity contours are more aligned with the line-of-sight (i.e. right panel in Figure \ref{fig_deproject1}).

The de-projected image in Figure \ref{fig_deproject1} show one 0.6 pc scale, elongated structure.
The ratio of the rotational period to the timescale of the global contraction can be approximated by $\frac{2\pi r}{v_{rot}}$/$\frac{r}{v_{in}}$$\sim$$\frac{2\pi}{0.422}$/$\frac{1}{0.906}$$\sim$6.74. 
This much longer rotational period than the timescale of the global contraction permits the existence of extended elongated structures. 

In Figure \ref{fig_deproject2}, we assume a rotationally dominant velocity field, where $\psi$=0$^{\circ}$ outside of the 0.16 pc radius, and $\psi$=15$^{\circ}$ inside of the 0.16 pc radius. 
Having this small value of $\psi$ inside of the 0.16 pc radius is to suppress the foreground/background degeneracies.
The left panel of Figure \ref{fig_deproject2} shows that this velocity model also reasonably explains the observed PV diagram, especially at outer than the $\sim$0.1 pc radius.
However, within the 0.1 pc radius, the velocity model shows tension with the observed PV diagram in the left panel of Figure \ref{fig_deproject2}.
The very narrow permitted velocity range given by the pure rotating velocity model can barely explain the significant emission at $\sim$$-$0.03 pc and $-$3 km\,s$^{-1}$, unless we artificially incorporate a few km\,s$^{-1}$ local linewidth into the velocity model.
Incorporating some infall velocity inside the 0.1-0.15 pc radius, and assuming a rotationally dominant motion at larger radii, may the most naturally explain the observed PV diagram, which was also suggested by the toy model presented in Section \ref{sec:toy}. 
Nevertheless, within the 0.1 pc radius, the de-projected image with such velocity model consistently shows protrusions, similar with the former case (see Figure \ref{fig_deproject1}).
The protrusions are connected with arm like features. 
On the larger scales, the de-projected structures appear relatively smooth.
The smoothness of the large-scale structures may be artificial, which can be either because of the intrinsic line width or our velocity resolution does not allow us to distinguish structures, or because our velocity model tends to give us this smoothed geometry. 
 
\begin{figure}
\vspace{-4.5cm}
\includegraphics[width=9cm]{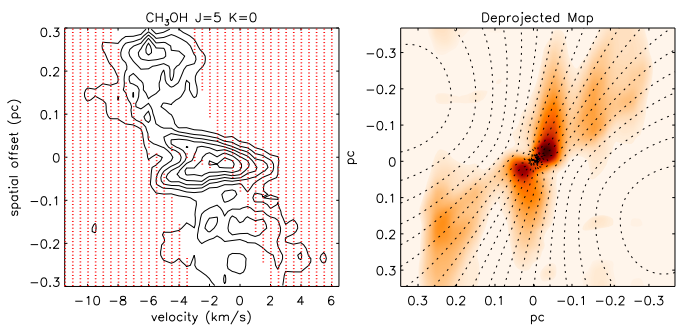}
\caption{\footnotesize{
Similar with Figure \ref{fig_deproject1}  ($\psi$ = 25$^{\circ}$). However, we adopt the minimal possible value of stellar mass of 69 $M_{\odot}$ suggested by Klaassen et al. (2009) and Keto, Zhang, \& Kurtz (2008).
}}
\label{fig_deproject3}
\end{figure}

\begin{figure}
\vspace{-4.5cm}
\includegraphics[width=9cm]{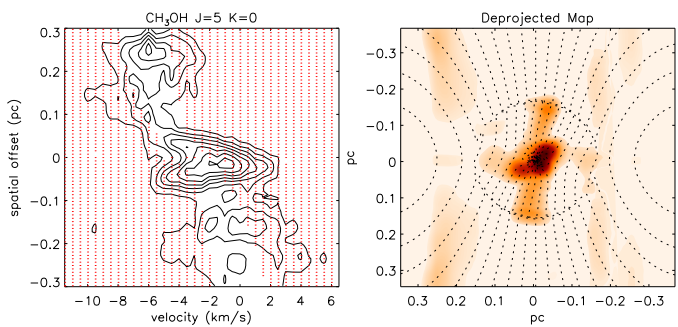}
\caption{\footnotesize{
Similar with Figure \ref{fig_deproject2} (i.e. rotational velocity dominated model; $\psi$ = 0$^{\circ}$ for $r$$>$0.16 pc; $\psi$ = 15$^{\circ}$ for $r$$<$0.16 pc). However, we adopt the minimal possible value of stellar mass of 69 $M_{\odot}$ suggested by Klaassen et al. (2009) and Keto, Zhang, \& Kurtz (2008).
}}
\label{fig_deproject4}
\end{figure}

Figure \ref{fig_deproject3} and \ref{fig_deproject4} show the results of de-projection, adopting the velocity model with the same assumption of $\psi$ with the velocity models adopted while generating Figure \ref{fig_deproject1} and \ref{fig_deproject2}.
However, in these two cases, we set the value of the centrally embedded stellar mass to be 69 $M_{\odot}$ (Klaassen et al. 2009), to demonstrate the effects of alternating the embedded mass.
The lower value of the embedded stellar mass sets the smaller scale of velocity in radius $\lesssim$0.1 pc. 
This pushes de-projected gas structures to be closer to the center. 
The de-projected image in Figure \ref{fig_deproject3} shows a dense molecular arc within the 0.1 pc region. 
The de-projected image in Figure \ref{fig_deproject4} shows a molecular arm wrapping the center within the 0.1 pc region. 
The de-projection of the structures at larger scale is less sensitive to the assumption of the centrally embedded stellar mass, since the molecular mass begins to dominate the total embedded mass.  

The deprojected images in Figure \ref{fig_deproject1}, \ref{fig_deproject2}, \ref{fig_deproject3}, and \ref{fig_deproject4} all suggest that inward of the $\lesssim$0.15 pc radius, the non-uniformity is essential.
Dense gas structures in this central region may be clumpy, or may consist of spiral arm-like structures.
The spiral arm-like structures may be extended to the outer radii.
Although the de-projected gas structures outside of the 0.15 pc radius appear different with different velocity assumptions (rotation+infall versus nearly pure rotation). 
We note that part of the difference may be related to our present way of treating the near/far side degeneracy.
For example, for the de-projection for the right panel in Figure \ref{fig_deproject2}, we can also choose to place all gas in between $0.15$-$0.3$ pc in the horizontal direction to the near side, and place all gas in between $-0.15$-$-0.3$ pc to the far side.
Then the de-projected structures on the extended spatial scale will appear very similar to those in Figure \ref{fig_deproject1}.
The de-projected, elongated gas arms (or filaments) may be connected with the $>$1 pc scale gas filaments resolved from the molecular line mapping observations on larger spatial scales (Liu et al. 2012a).
The higher angular and velocity resolution observations can further alleviate some elongation artifacts in de-projection (more in Section \ref{sub:caveats}).

\section{Enclosed mass and Toomre's instability}\label{sec:toomre}
To understand the origin of azimuthal asymmetry in G10.6-0.4, we examine the approximated self-gravitational instability by evaluating the Toomre Q parameter as a function of radius, assuming a flattened rotating disk.
The Toomre Q parameter is defined as:
\begin{equation}
Q \equiv \frac{C_{s}\kappa }{\pi G\Sigma},
\end{equation}
where $C_{s}$ is the sound speed of the molecular gas, $\kappa$ is the epicyclic frequency, G is the gravitational constant, and $\Sigma$ is the disk surface mass density.
The numerical hydrodynamical simulations (Nelson et al. 1998) suggests that the spiral arms start forming when the minimum value of Q in the disk is around 1.5--2.
When Q$<$1, the disk is unstable under self-gravity for all wave number $k$, and we can expect a clumpy morphology.

\begin{figure}
\hspace{-1cm}
\includegraphics[trim={0 0 0 0.65cm}, clip, width=10cm]{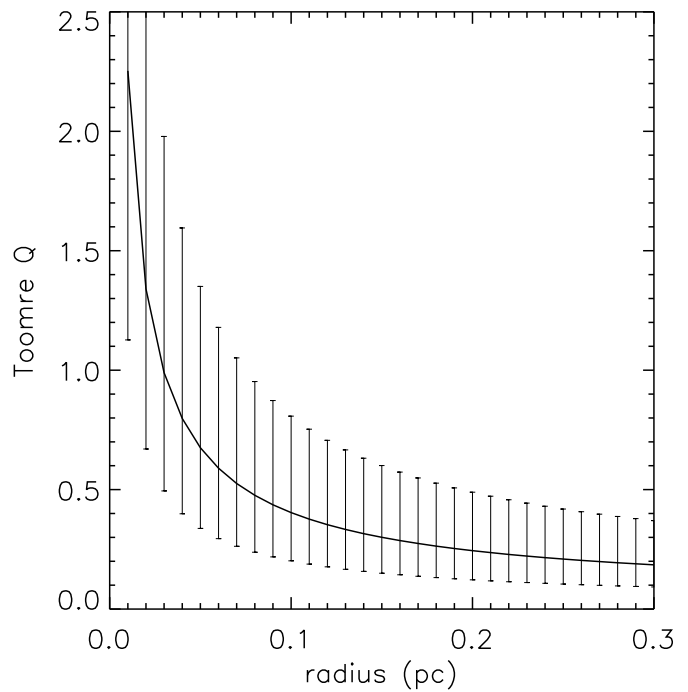}
\vspace{-0.5cm}
\caption{
\footnotesize{Toomre Q parameter as a function of radius, for G10.6-0.4.
}
}
\label{fig:toomreQ}
\end{figure}

When deriving $\kappa$, we adopt the velocity curve described by Equation \ref{eq_v}.
In addition, we replace $C_{s}$ with an assumed turbulence velocity on the 0.3 pc scale\footnote{The typical turbulence velocity in molecular clouds follow the scaling law $\Delta v\approx \mbox{1 km s}^{-1} (\mbox{R/[1 pc]})^{\beta}$, with $\beta\approx$0.38-0.5. The value of $\Delta v$ is about 0.6 km\,s$^{-1}$ at the 0.3 pc scale. The $\Delta v$ is constrained to be $\lesssim$1 km\,s$^{-1}$ in G33.92+0.11 Main (Liu et al. 2012).}.
For simplicity, we assume that the enclosed gas mass is smoothly and axi-symmetrically distributed.
The evaluated Toomre Q parameter as a function of radius is presented in Figure \ref{fig:toomreQ}.
The results suggest that the accretion flow is generally Toomre unstable on the resolved spatial scales (0.05-0.3 pc radius), which can explain the existence of spiral arms and localized dense gas condensations.
We may underestimated $\kappa$ inside the $\sim$0.05 pc radius, since the highly gravitationally accelerated gas may be largely photo-ionized.
Whether or not the accreting gas on such a small spatial scale is Toomre stable is not very well constrained by the existing observations.
On the $>$0.05 pc scale, the gas accretion flow may be more gravitationally unstable than how is indicated by our present estimates of Toomre Q parameter, since the gas is likely concentrated to azimuthally asymmetric dense structures rather than being smoothly distributed (Section \ref{sec:toy}, \ref{sec:deproject}).

\section{Discussion}\label{sec:discussion}
\subsection{An overall picture for the inner 0.3 pc radius of G10.6-0.4}\label{sub:g10}
The overall morphology of accretion flow within the 0.3 pc radius from the central OB cluster of G10.6-0.4, may resemble a gravitational unstable, flattened rotating disk, where the highest mass stars are located at the center of the system (see Figure \ref{fig:schematic} for a schematic picture).
In this system, molecular gas mass dominates the overall enclosed mass, and the accretion flow becomes gravitationally unstable.
Large (0.1-0.3 pc) scale spiral arm-like structures form, which may subsequently fragment into localize dense gas cores or condensations.
Individual of these gas cores or condensations may form intermediate- or low-mass stars, which can emanate molecular outflows our jets (Liu et al. 2011).

The gravitational instability in this accretion flow is consistent with the detection of multiple UC H\textsc{ii} region within the mid-plane of it (e.g. the satelliate OB clusters in Figure \ref{fig:overview}B; see also Ho \& Haschick 1986; Liu et al. 2011).
In addition, the very high angular resolution (0$\farcs$1) resolution of the 1.3 cm free-free continuum emission have resolved several ionized arcs eastern of the central OB cluster (Sollins \& Ho 2005).
The fact that the ionized arcs were only observed from the east of the central OB cluster implies that (1) there is a molecular gas gap in between the central OB cluster and the ionized arcs, such that the ionized arc can be illuminated from the central OB cluster, and (2) the western side from the central OB cluster is either very well shielded by dense gas, or may have very different gas morphology from the eastern gas structures.
These are consistent with the highly azimuthally asymmetric molecular gas structures around the central OB cluster.
The asymmetric gas structures, or the spiral arm-like structures, may help redistribute the angular momentum, which may consistently explain the observed infall velocity (Keto et al. 1987; Keto 1990; present work), and the rapidly decreasing specific angular momentum from outer to inner radii (Liu et al. 2010a).

We note that in the de-projection, not all assumptions of velocity field and enclosed stellar mass which can explain the observed PV diagrams, are physically plausible.
For example, assuming a 175 $M_{\odot}$ enclosed stellar mass and less dominant rotational motion (i.e. $\psi$$>$25$^{\circ}$) will imply that the blue and redshifted gas components within the $\pm$0.1 pc offset positions, are more separated from the central OB cluster than how they are presented in Figure \ref{fig_deproject1}.
The de-projected structures within the $\pm$0.1 pc offset with such assumptions will become more (artificially) elongated, or even separated in the line-of-sight, which cannot confine the central UC H\textsc{ii} region.
Increasing the enclosed stellar mass for several times will significantly enlarge the permitted line-of-sight velocity range for explaining the observed PV diagrams, and will increase the Toomre Q parameter such that the system appear (more) stable.
However, the increased stellar mass will also imply the further de-projected line-of-sight distances of gas components from the central OB cluster, which is not plausible. 
The enclosed stellar mass at the center of G10.6-0.4 therefore cannot be very different from the range of 69-175 $M_\odot$, to allow sufficient permitted velocity range for explaining the observed PV diagram while avoiding the artificial elongation in the line-of-sight.

The dense gas morphology around G10.6-0.4 we derived may be plausible.
We note that the previous SMA, VLA, and Atacama Large Millimeter Array (ALMA) studies on the similar system G33.92+0.11 have spatially resolved dense, spiral arm-likes gas structures and many localized dense cores, which are orbiting the centralized UC H\textsc{ii} region (Liu et al. 2012b; Liu et al. 2015).
In addition, in G33.92+0.11, the resolved localized dense cores appear to be self-gravitating and are forming stars, which were diagnosed from the high velocity SiO 5-4 jets (Minh et al. 2016).
A dust (and gas) spiral were also resolved within the $\sim$0.2 pc radius around the OB cluster-forming region NGC 7538 IRS. 1, although it is not yet certain whether this spiral arm-like structure was formed due to gravitational instability, or was due to the interaction with molecular outflows (Wright et al. 2014). 

The spatial non-uniformity of the gravitational unstable accretion flow may directly link to the time variability of accretion rate, and the variations of the radio fluxes and sizes of the UC H\textsc{ii} region (e.g. Galv\'an-Madrid et al. 2009).
It may be similar to a scaled-up model of episodic low-mass (proto)stellar accretion via gravitational unstable disk (e.g. Vorobyov \& Basu 2015; Sakurai et al. 2016).

\subsection{Physics and applications}\label{sub:physics}
Important applications of the proposed analysis techniques include the observations of the inner few thousand AU accretion flows around the young OB star or star cluster, and the observations of the inner few tense AU of the Class 0/I, or FU Orionis disks (e.g. Dong et al. 2016; Liu et al. 2016).
For low-mass Class 0/I objects, these observations are the key to address the role of gravitational instability in assisting the (proto)stellar accretion (Vorobyov et al. 2015; Klassen et al. 2016).

\begin{figure}
\hspace{0.3cm}
\includegraphics[width=8.5cm]{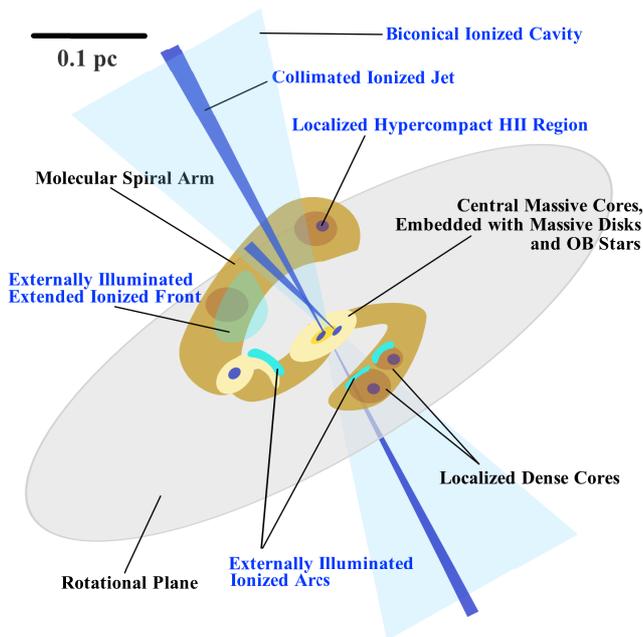}
\caption{\footnotesize{
Schematic picture for the distribution of dense molecular and ionized gas in a gravitationally unstable, OB cluster-forming molecular clump. Un-shield molecular gas condensations can be externally illuminated and show varieties of ionized features. On the other hand, bright point-like emission sources are expected to be embedded by dense molecular lumps. In low angular resolution observations, this variety of structures will be smeared into an integrated UC H\textsc{ii} region.
}}
\label{fig:schematic}
\end{figure}

For the young and massive OB clusters, the observations of how localized dense gas cores/condensations form in the gravitational unstable gas converging flow, and the observations of the detailed gas kinematics, may provide clues for how to realize the high-mass end of the stellar initial mass function (IMF).
In particular, the observations of Larson (1982) have suggested a correlation between the maximum stellar mass in a cluster ($m_{\mbox{\scriptsize{max}}}$) and the mass of the natal molecular cloud ($M_{\mbox{\scriptsize{cloud}}}$), such that $m_{\mbox{\scriptsize{max}}}$=0.33\,$M_{\mbox{\scriptsize{cloud}}}^{0.43}$.
Based on statistical analyses on the observed $m_{\mbox{\scriptsize{max}}}$ from a sample of young (age$<$4 Myr) stellar clusters, Kroupa \& Weidner (2005), Weidner \& Kroupa (2006), and Weidner et al. (2010) demonstrated a tight correlation between $m_{\mbox{\scriptsize{max}}}$ and the cluster mass $M_{\mbox{\scriptsize{ecl}}}$, which indicated that $m_{\mbox{\scriptsize{max}}}$ of $>$100 $M_{\odot}$ stellar clusters cannot be represented by random samplings of the canonical stellar IMF.
Especially for the $>$100 $M_{\odot}$ stellar clusters, the masses of their highest mass stars may be determined by rather deterministic and trackable physical processes.
Weidner \& Kroupa (2006) proposed that star clusters form in an ordered fashion, starting with the lowest-mass stars until feedback can outweigh the gravitationally induced formation process.
An alternative interpretation may be that the highest mass one or few stars form in relatively special environments of the natal molecular cloud.
For the cases of OB cluster-forming regions like G10.6-0.4 or G33.92+0.11 (Liu et al. 2015), this environment is the most likely to be the $<$1 pc scale flatten accretion flows deeply embedded at their centers.
The masses of the highest mass star(s) are thereby determined by the accretion processes and fragmentation within such environment.
Taking Figure \ref{fig:schematic} for instance, the (first) highest mass star may only form in the unique {\it central massive core}, which with the aid of the global contraction, can quench the UC H\textsc{ii} region for longer time.
The {\it satellite cores} are also massive, and can form satellite OB clusters and localized UC H\textsc{ii} regions (e.g. Figure \ref{fig:overview}B).
However, the satellite cores are more easily dispersed by stellar feedback so will form less massive stars than the central molecular core.
To begin our understanding of the highest mass end of the stellar IMF in massive clusters, and the understanding of the correlation between $m_{\mbox{\scriptsize{max}}}$ and $M_{\mbox{\scriptsize{ecl}}}$ and $M_{\mbox{\scriptsize{cloud}}}$, it is therefore very important to increase the number of case studies which the details of density structures and kinematics are resolved.

The known gravitational unstable disk-like objects are rare, and therefore is hard to obtain a sample of face-on sources.
In addition, they are likely to be embedded by exterior dense gas structures, such that the observations of face-on sources are not necessarily easy to interpret, due to blending.
Some inclination of the source helps differentiate the gravitationally accelerated gas structures in the inner region from the exterior gas structures, from the velocity domain. 
However, most of these sources are relatively distance, such that even the Atacama Large Millimeter Array (ALMA) observations with a few hours of exposure time would only marginally resolve the gas structures.
For some sources, it may eventually require to diagnose azimuthal asymmetry from the velocity domain.

\subsection{Uncertainty, caveats, and ways out}\label{sub:caveats}
We have provided examples of interpreting the asymmetric PV diagrams based on modeling and direct de-projection.
Simple models can provide a good intuition about the morphology of the target source. 
For the target sources which consist of complicated structures, modeling may however lose objectiveness. 
Direct de-projection based on a velocity model may systematically reconstruct structures of the target sources.
However, the results of de-projection still require careful examination and interpretation.
In particular, for a perfectly edge-on source with pure rotational gas motions,  de-projection is seriously subject to the near and far side degeneracy.
Moreover, in the de-projected image, gas structures with broad linewidth may become artificially elongated in the spatial direction parallel with the line-of-sight.
These problems are fundamental, similar to the ambiguity faced when deriving the spiral arm structures of the Milky Way based on CO line surveys.

For the (marginally) resolved, geometrically thin gas structures (e.g. some protoplanetary disks), de-projecting the PPV image cubes instead of the PV diagrams to the position-position-position (PPP) space may greatly suppress the degeneracy.
The observations of multiple gas temperature tracers may further help locate gas structures from the dominant heating sources. 
To avoid artificially stretching gas structures in the direction of the line-of-sight, we may de-compose the spatially compact and extended structures in the PV diagram before de-projecting (e.g. applying high-/low-pass filters or performing wavelet transform), and then require the de-projected compact gas structures to have similar or identical sizescales in the line-of-sight and the perpendicular dimensions.
Practically how this can be done depends also on the spatial and spectral resolutions of the observations, which may require some tests utilizing synthesized images from numerical simulations, which is beyond the scope of the present paper.

Finally, either modeling or de-projection require assumptions about the velocity field.
When the gravitational force is dominated by a centralized compact source (e.g. a (proto)star or a high-mass molecular core), it may be fair to assume that the velocity field is axi-symmetric.
Without a dominant centralized gravitational source, the velocity field may still be approximately axi-symmetric if the is sufficient relaxation of gas motions (e.g. the case of spiral galaxies).
Observations of very complicated system (e.g. binary, triple, or multiple systems) should be interpreted with more care. 

\begin{acknowledgements}
The Submillimeter Array is a joint project between the Smithsonian
Astrophysical Observatory and the Academia Sinica Institute of Astronomy
and Astrophysics, and is funded by the Smithsonian Institution and the
Academia Sinica (Ho et al. 2004).
The National Radio Astronomy Observatory is a facility of the National
Science Foundation operated under cooperative agreement by Associated
Universities, Inc.
HBL thanks Dr. Pavel Kroupa for an informative discussion.
HBL thank the referee for instructive suggestions.
\end{acknowledgements}


\end{document}